\documentclass[12pt,english,american,onecolumn]{article} 
\usepackage{bm}%
\usepackage[normalem]{ulem}
\usepackage[colorlinks=true,linkcolor=blue,citecolor=black]{hyperref}%
\usepackage{color}%
\usepackage[T1]{fontenc}
\usepackage[utf8]{inputenc}
\usepackage{algorithm}
\usepackage[font={small}]{caption}
\usepackage{authblk}
\usepackage{float}
\usepackage{mathtools}
\usepackage{amsmath}
\usepackage{amsthm}
\usepackage{amssymb}
\usepackage{graphicx}
\usepackage{color}
\usepackage{graphicx,epstopdf}
\usepackage{algpseudocode}
\usepackage[super]{natbib}
\usepackage{enumerate}
\usepackage{etoolbox}
\usepackage{lmodern}
\usepackage{stfloats}
\usepackage{subfig}

\makeatother

\makeatletter
\def\blfootnote{\xdef\@thefnmark{}\@footnotetext}
\makeatother

\floatstyle{ruled}
\newfloat{algorithm}{tbp}{loa}
\providecommand{\algorithmname}{Algorithm}
\floatname{algorithm}{\protect\algorithmname}

\expandafter\ifx\csname package@font\endcsname\relax\else
 \expandafter\expandafter
 \expandafter\usepackage
 \expandafter\expandafter
 \expandafter{\csname package@font\endcsname}%

\fi
\author[1,4]{Mattia Serra\thanks{serram@ucsd.edu}}
\affil[1]{\footnotesize University of California San Diego, Department of Physics, CA 92093, USA}%
\author[2]{Linnea Lemma}
\affil[2]{University of California at Santa Barbara, Santa Barbara, CA 93111, USA}
\author[3]{Luca Giomi}
\affil[3]{Instituut-Lorentz, Universiteit Leiden, P.O. Box 9506, 2300 RA Leiden, Netherlands}
\author[2]{Zvonimir Dogic}
\author[4,5,6]{L. Mahadevan\thanks{lmahadev@g.harvard.edu}}
\affil[4]{Paulson School of Engineering and Applied Sciences, Harvard University, Cambridge, MA 02138, USA}
\affil[5]{Department of Organismic and Evolutionary Biology, Harvard University, Cambridge, MA 02138, USA}
\affil[6]{Department of Physics, Harvard University, Cambridge, MA 02138, USA}

\title{Defect-mediated dynamics of coherent structures in active nematics}%

\definecolor{red}{rgb}{0.75,0,0}
\definecolor{blue}{rgb}{0,0,0.75}
\definecolor{green}{rgb}{0,0.5,0}
\definecolor{black}{rgb}{0,0,0}

\newcommand{\black}[1]{{\color{black} #1}}

\begin{document}

\maketitle

\begin{abstract}
Active fluids, such as cytoskeletal filaments, bacterial colonies and epithelial cell layers, exhibit distinctive orientational coherence, often characterized by nematic order and topological defects. By contrast, little is known about positional coherence --  i.e., how a hidden dynamic skeleton organizes the underlying chaotic motion -- despite this being one of their most prominent and experimentally accessible features.
Using a combination of dynamical systems theory, experiments on two-dimensional mixtures of microtubules and kinesin and hydrodynamic simulations, we characterize positional coherence in active nematics. These coherent structures can be identified in the framework of Lagrangian dynamics as moving attractors and repellers, which orchestrate complex motion. To understand the interaction of positional and orientational coherence on the dynamics of defects, we then analysed observations and simulations and see that +1/2 defects move and deform 
the attractors, thus functioning as control centers for collective motion.
Additionally, we find that regions around isolated +1/2 defects undergo high bending and low stretching/shearing deformations, consistent with the local stress distribution. The stress is minimum at the defect, while high differential stress along the defect orientation induces folding. Our work offers a new perspective to describe self-organization in active fluids, with potential applications to multicellular systems. 
\end{abstract}

\section*{Introduction}

\noindent \color{black} 
Many
\black{out-of-equilibrium systems}, \black{from bird flocks down to biofilms and the cell cytoskeleton,} consist of agents that consume energy and self-organize into large-scale patterns and collectively moving structures 
\cite{Vicsek2012,RevModPhys.85.1143,kruse2004asters,ballerini2008interaction,zhang2010collective,bricard2013emergence,dombrowski2004self}, that are breathtaking in their beauty and complexity  while being of relevance to questions of embryonic development, wound healing and cancer \cite{friedl2009collective,ladoux2017mechanobiology,Doostmohammadi2018,serra2020dynamic}. Understanding the mechanisms that lead to these patterns and characterizing the phases of active matter systems will unravel their complexity, while also suggesting ways to mimic them using synthetic materials, and eventually control and design active systems using external fields. 

Self-organized patterns are typically described in the language of Eulerian \black{flows, where the local velocity, pressure and orientation of the active building blocks are treated as fields within a fixed control volume. Two-point correlation functions, spectral densities and other quantities} inspired by studies of statistical steady states of turbulence in Newtonian fluids, \black{are typical outcomes of this approach}  \cite{wensink2012meso,Giomi2015}. 
\black{Topological defects, i.e. localized singularities in the orientation of the active building blocks \cite{RevModPhys.85.1143},} have \black{also} been extensively studied and their dynamics is known to \black{be inherently entangled with the large-scale chaotic flow} 
\cite{Giomi2015,PhysRevX.9.041047,Tan2019}. But what is the relation between the dynamics of defects and the large scale coherence in the flow fields that are typically spatially heterogeneous and temporally unsteady?

A natural framework to address this question is provided by the Lagrangian description of fluid flow. By tracing the motion of passive particles in unsteady flows, which may also include chaotic paths, one can often identify robust skeletons, commonly referred to as Coherent Structures (CSs) \cite{LCSHallerAnnRev2015,SerraHaller2015,hadjighasem2017critical}, which shape trajectory patterns and reveal the organizing barriers to material transport (see e.g. Refs. \cite{serra2017uncovering,serra2020SR}). Here, we combine theoretical concepts from nonlinear dynamics, active hydrodynamics simulations and experiments on suspensions of microtubules and kinesin to unravel the CSs underlying the chaotic flow of two-dimensional active nematics and their relation to the dynamics of topological defects. 

\section*{Results}

\subsection*{Lagrangian deformations and coherent structures}
\begin{figure}[h!]
	\centering
	\includegraphics[height=0.5\columnwidth]{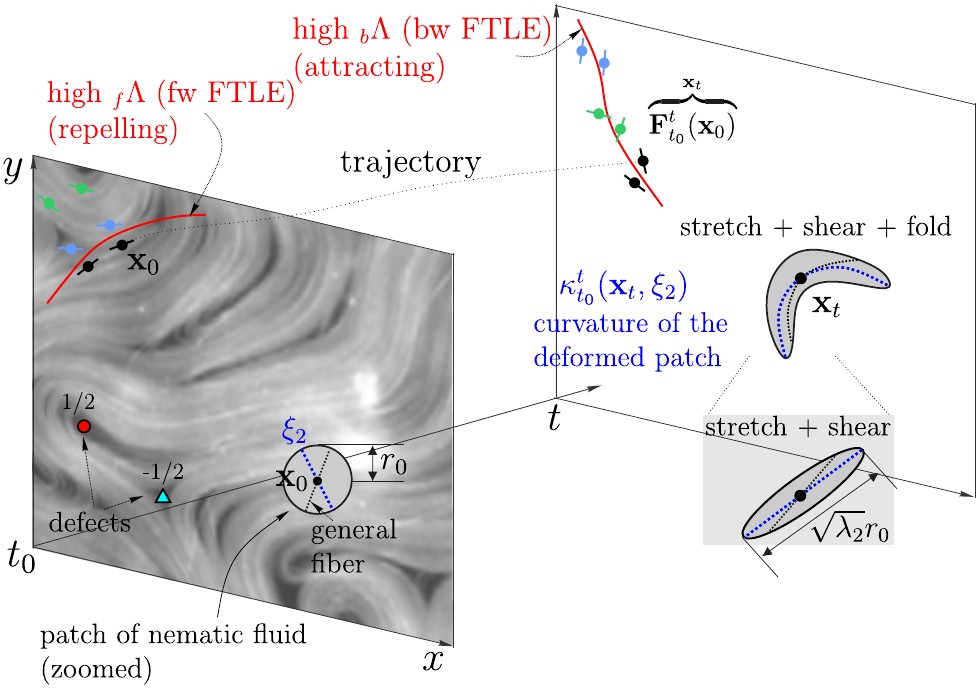}
	\caption{Lagrangian view of an active nematic fluid. Red curves demarcate repelling and attracting regions in the flow identified by high foreword and backward FTLE values. Repellers are based at the initial fluid configuration while attractors at the final one. Initially close tracers that are at opposite sides of a fw FTLE ($_f\Lambda$) ridge will move far apart at time $t$. Similarly, initially distant tracers are attracted to a bw FTLE ($_b\Lambda$) ridge at time $t$. An infinitesimal patch of nematic fluid at $\mathbf{x}_0$ will get stretched and folded over $[t_0,t]$. Different fibers in this patch tend to align to a distinguished fiber $\xi_2(\mathbf{x}_0)$ during $[t_0,t]$, and the curvature of the folded patch is $\kappa_{t_{0}}^t(\mathbf{x}_t,\mathbf{\xi}_2)$.}	
	\label{fig:FTLEScheme}
\end{figure}

Our theoretical and computational framework for kinematic analysis starts by considering the velocity field $\mathbf{v}(\mathbf{x},t)$ of a planar active nematic fluid, and the corresponding flow map 
\begin{eqnarray}
\mathbf{F}_{t_{0}}^t(\mathbf{x}_0)=\mathbf{x}_0+\int_{t_{0}}^{t}\mathbf{v}(\mathbf{F}_{t_{0}}^\tau(\mathbf{x}_0),\tau)\ d\tau\;. 
\label{eq:Lagrmaps}
\end{eqnarray}
This evolves \black{the} initial position $\mathbf{x}_0$ \black{of a virtual tracer particle} to the corresponding position $\mathbf{F}_{t_{0}}^t(\mathbf{x}_0)$ \black{at time $t$}, \black{whereas} the spatial heterogeneity of $\mathbf{F}_{t_{0}}^t(\mathbf{x}_0)$ describes Lagrangian deformation of the nematic fluid over $[t_0,t]$. Locally, a small fluid patch can be stretched, sheared and folded. Stretching and shearing (Fig. \ref{fig:FTLEScheme}) are completely characterized by the right Cauchy-Green strain tensor field $\mathbf{C}_{t_{0}}^{t}(\mathbf{x}_0)$ \cite{TruesdellNoll2004} defined as
\begin{equation}
\black{\mathbf{C}_{t_{0}}^{t}(\mathbf{x}_0)= \mathbf{\nabla F}_{t_{0}}^{t}(\mathbf{x}_0)^{\top}\cdot\mathbf{\nabla F}_{t_{0}}^{t}(\mathbf{x}_0)\;,}\label{eq:CG}
\end{equation}
where $\mathbf{\nabla F}_{t_{0}}^{t}(\mathbf{x}_0)$ is the Jacobian of the flow map. 

\black{Given a material patch consisting of straight {\em fibers} at the initial time (Fig. \ref{fig:FTLEScheme}), and} denoting by $\lambda_1\leq\lambda_2$ and \black{$\{\bm{\xi}_1, \bm{\xi}_2\}$} the eigenvalues and the associated \black{orthonormal} eigenvectors of $\mathbf{C}_{t_{0}}^{t}(\mathbf{x}_0)$, \black{one can} interpret $\bm{\xi}_2$ \black{as the} fiber\black{'s} most stretched \black{direction} (by a factor $\sqrt{\lambda_2}$) \black{and} $\bm{\xi}_1$ \black{as} the least stretched one (by a factor $\sqrt{\lambda_1}$). 
In chaotic systems $\lambda_2$ usually grows exponentially in time, i.e. $ \lambda_2 \sim e^{\sigma t}$, and it is typically rescaled as
\begin{equation}
\Lambda_{t_0}^t(\mathbf{x}_0)= \frac{1}{\vert t-t_0 \vert}\log \sqrt{\lambda_2(\mathbf{x}_0)},
\label{eq:FTLE}
\end{equation} 
which denotes the largest Finite-Time Lyapunov Exponent (FTLE).
The maximal stretching along $\bm{\xi}_2$ implies that, over time, other fibers align with the local $\bm{\xi}_2$ direction, as illustrated in Fig. \ref{fig:FTLEScheme}. This alignment property follows from the corresponding asymptotic result \cite{Giona1998}.

\black{By contrast,} folding deformations \black{determine variations} in the curvature of material fibers over time \cite{Serra2017Separation}. \black{This can be computed from the} second order spatial derivatives of $\mathbf{F}_{t_{0}}^t(\mathbf{x}_0)$ \black{and provides} additional information relative to stretching and shearing. 
Starting with an infinitesimal patch of straight fibers at $\mathbf{x}_0$, from the $\bm{\xi}_2$-alignment property follows that the most observable fluid folding -- i.e. what a colored patch would show -- during $[t_0,t]$ is the one along the $\bm{\xi}_2$ fiber (Fig. \ref{fig:FTLEScheme}). We compute this folding as
\begin{equation}
\black{\kappa_{t_{0}}^{t}({\bf x}_{0},\bm{\xi}_{2}) = \frac{[(\mathbf{\nabla}^2 \mathbf{F}_{t_0}^t(\mathbf{x}_0)\bm{\xi}_2)\cdot\bm{\xi}_{2}]\cdot[\mathbf{\nabla_{\perp} F}_{t_0}^t(\mathbf{x}_0)\cdot\bm{\xi}_2]}{\lambda_{2}^{3/2}}\;,}
\label{eq:Kappa}
\end{equation} 
where $(\nabla^2 F_{t_{0}}^t(\mathbf{x}_0)\mathbf{\xi}_2)_{ij}=\tiny{\sum\limits_{k}}{F_{t_{0}}^t}_{i,jk}(\mathbf{x}_0)\mathbf{\xi}_{2_k},\ i,j,k\in\{1,2\}$ and  \black{$\nabla_{\perp}=(-\partial_{y},\partial_{x})$}, (See Supplementary Information - SI, Section 1). 
\black{The curvature} $\kappa_{t_{0}}^t(\mathbf{x}_0,\bm{\xi}_2)$ \black{displays} the Lagrangian folding that occurred from $t_0$ to $t$ \black{over} the initial fluid configuration. \black{Analogously, the same folding can be displayed over the final position by} transporting $\kappa$ along trajectories (Fig. \ref{fig:FTLEScheme}), \black{namely:}
\begin{equation}
\kappa_{t_{0}}^t(\mathbf{x}_t,\bm{\xi}_2)=\kappa_{t_{0}}^t(\mathbf{F}_{t_0}^t(\mathbf{x}_0),\bm{\xi}_2)\;.
\label{eq:Kappaft}
\end{equation} 
Large values of $\kappa_{t_{0}}^t(\mathbf{x}_0,\bm{\xi}_2)$ \black{mark} the initial positions $\mathbf{x}_0$ of the nematic fluid that will undergo large \black{folding} in the \black{time} interval $[t_0,t]$, and, similarly  $\kappa_{t_{0}}^t(\mathbf{x}_t,\bm{\xi}_2)$ identifies the final positions of nematic fluid that experienced small/large curvature changes. In the SI Section 1, we provide general expressions of the material curvature and its alternative formulation in terms of Eulerian quantities such as flow vorticity, divergence and rate-of-strain tensor. Altogether, Eqs. (\ref{eq:FTLE}-\ref{eq:Kappaft}) completely quantify the maximal stretching and folding deformations of a continuum moving under a given flow map $\mathbf{F}_{t_{0}}^{t}$. These measures are model-independent (or agnostic to the mechanisms driving the flow), hence applicable to experimental or computational velocity fields and \black{readily} implement\black{able}. 

The forward FTLE (fw FTLE or $_f\Lambda$), is a scalar field over the initial particle positions $\mathbf{x}_0$ that quantifies the maximum local deformation and identifies the location of maximum spatial separation of initially close particles over the time interval $[t_0, t]$. Similarly, the backward FTLE (bw FTLE or $_b\Lambda$), defined over the final positions $\mathbf{x}_t$, identifies the location of maximum spatial convergence of initially distant particles over $[t_0, t]$. Together they demarcate regions of attraction (attracting CS) and repulsion (repelling CS) (Fig. \ref{fig:FTLEScheme}), but do not distinguish whether attraction or repulsion are \black{caused by} shear or normal deformations \cite{LCSHallerAnnRev2015,SerraHaller2015}.

\subsection*{Coherent structures organize particle motion}
We now \black{employ} these analytical tools to analyze 2D microtubule based active nematic liquid crystals assembled on a surfactant stabilized oil-water interface \cite{Sanchez2012}. Large scale chaotic dynamics of these materials is collectively driven by kinesin molecular motors that move along multiple filaments to induce relative filaments sliding. 

\begin{figure*}[h!]
	\centering
	\includegraphics[width=1.0\columnwidth]{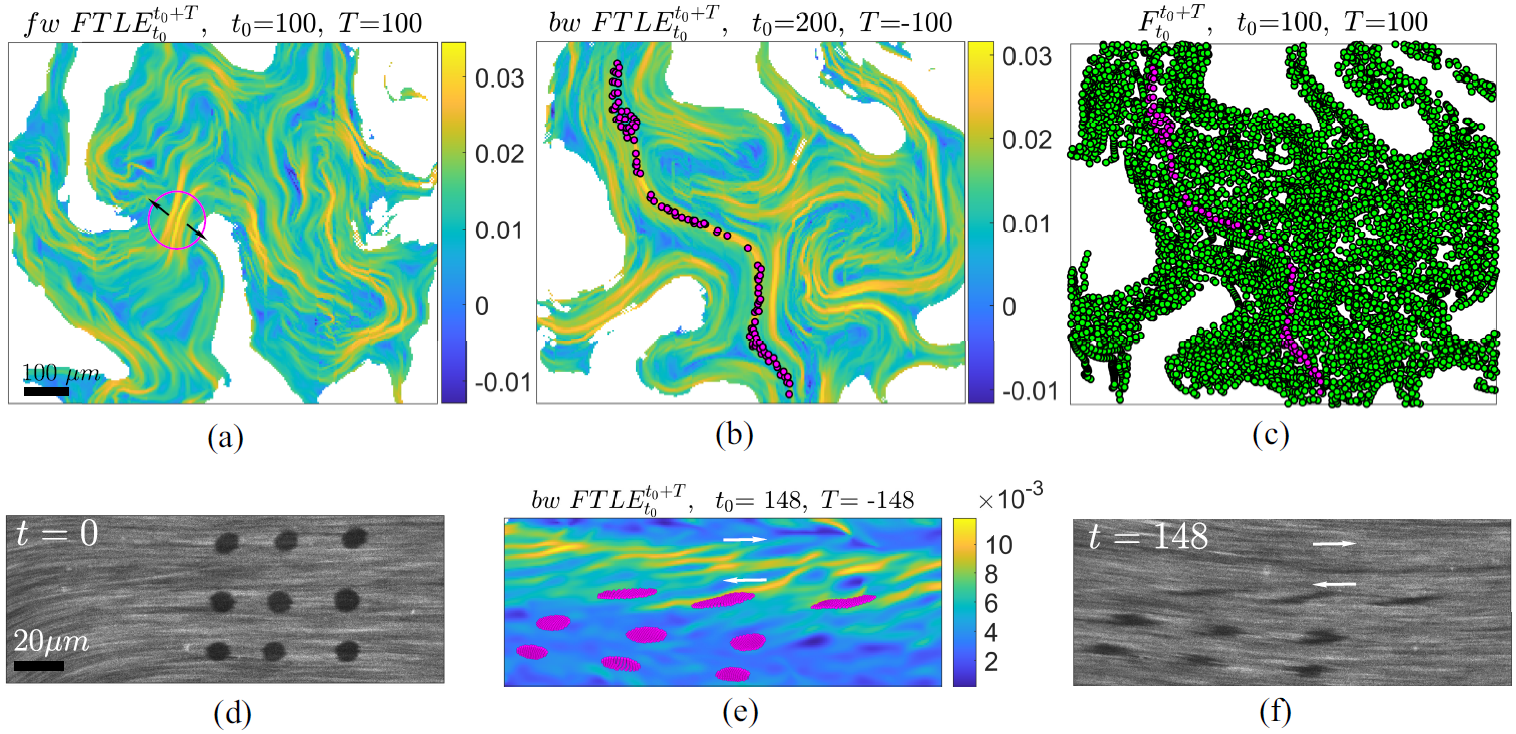}
	\caption{Dynamics of 2D microtubule based active nematics assembled on the oil-water interface. The PIV velocity is reconstructed on a uniform $670 \times 800\,\mu$m grid with \black{spatial resolution} $15.6\,\mu$m and \black{temporal resolution of} $1$ frame per second. (a) $_f\Lambda$ field whose ridges signal repelling CSs. White regions demarcate the set of particles that left the domain where the velocity field is available. (b) $_b\Lambda$ field whose ridges signal attracting CSs. Magenta dots represent the final ($t = 200$) position of nematogens started inside the magenta circle  shown in panel a at the initial time $t = 100$. (c) Final position of nematogens started from a uniform grid at the initial time. Particles started outside the magenta circle are  green. The time evolution is available as \href{https://www.dropbox.com/s/eljciu5cy49kqts/MovieNematoExp250uMATP_T100t0_100_DensetrajSpecLoc_v5.mp4?dl=0}{Movie1}. (d-f) Fluorescence recovery after photobleaching experiment in an active nematics. (d,f) Initial and final configuration of the FRAP experiment. (e) $ _b \Lambda _{148}^0$ along with advected particles at $t=148$, initialized in correspondence of the photobleached regions at $t=0$. \href{https://www.dropbox.com/s/ewxmm9ntnv8l6ep/MovieNematoExpFRAPseries28_T50_DensetrajSpecLoc_FltVel_SpInt_vBWFTLE_FRAP_res3.mp4?dl=0}{Movie2} shows panels d,e for increasing $t$. Time is \black{expressed} in \black{seconds (s)}. Colorbars encode attraction or repulsion rates in s$^{-1}$. In panels a-c, the ATP concentration is 250 $\mu$M. In panels d-f, the ATP concentration is 18 $\mu$M. 
	}
	\label{fig:FTLE250uMATP}
\end{figure*}

\black{Using Particle Image Velocimetry (PIV), we reconstruct} the velocity field of autonomously flowing active nematics. From the velocity field, we compute $_f\Lambda$ and $_b\Lambda$ for different time scales $\vert T\vert = \vert t-t_0 \vert$ using Eqs. (\ref{eq:Lagrmaps}-\ref{eq:FTLE}). Figures \ref{fig:FTLE250uMATP}a,b show the $_f\Lambda$ and $_b\Lambda$ fields for $\vert T\vert = 100s$. 
As sketched in Fig. \ref{fig:FTLEScheme}, particles \black{are} repelled from a $_f\Lambda$ ridge and attracted towards a $_b\Lambda$ ridge (covered by magenta dots). Panel c shows the final position of a set of particles initially released from a uniform grid and serves as a tracker of particles motion. 
\href{https://www.dropbox.com/s/qj50bcafqil0i3h/NematoExp250uMATP_FTLE.mp4?dl=0}{Movie1} shows the time evolution of the FTLE fields and particle positions. Although fluid moves chaotically, there is an underlying hidden coherent skeleton, captured by the FTLE fields, that dynamically organize their motion while remaining invisible to the inspection of fluid tracers. 

The FTLE also provides a \black{$|T|-$}dependent map of the stretching and shearing Lagrangian deformation of the active continuum, with ridges that demarcate sets of the fluid that will experience a distinguished deformation relative to their neighbors. This Lagrangian deformation consistently integrates along trajectories the separate contributions of viscous, elastic and active stresses deforming the nematic fluid, and thus encodes a memory trace of the nematodynamic field.

To correlate these computed Lagrangian memory traces with direct observations of the deformation patterns near $_b\Lambda$ ridges, we label regions of microtubule based active nematics and observe their subsequent evolution (SI, Section 5). \black{This is achieved by} photobleach\black{ing} nine circular regions of radius $\approx 4 \mu m$ (Fig. \ref{fig:FTLE250uMATP}d).  Using PIV data, we compute the $_b\Lambda$ field along with the position of Lagrangian tracers (magenta) initialized at $t=0$ in correspondence of the photobleached regions (\black{F}ig. \ref{fig:FTLE250uMATP}e). Stripe-shaped ridges of $_b\Lambda_{148}^0$  reveal a horizontal shear layer, along with  regions of distinctly high attraction and Lagrangian deformations. Our analysis predicts the evolution of advected and diffused photobleached patches (Fig. \ref{fig:FTLE250uMATP}f,  \href{https://www.dropbox.com/s/ewxmm9ntnv8l6ep/MovieNematoExpFRAPseries28_T50_DensetrajSpecLoc_FltVel_SpInt_vBWFTLE_FRAP_res3.mp4?dl=0}{Movie2}). Overall, $_b\Lambda_{T}^0$ provides a $\vert T \vert -$dependent map of attraction and stretch or shear deformations over the entire domain. 

\subsection*{Positive defects mediate attracting coherent structures}
Having uncovered the invisible organizers of the flow fields using Lagrangian coherent structures, we now turn to understand if and how they are related to the visible dynamics of topological defects, well known to be correlated with complex, large-scale nematodynamic flows \cite{Giomi2015,Tan2019}. As previously, we use microtubule based 2D active nematics powered by kinesin motors. By assembling slow nematics at 2 $\mu$M ATP concentration, we simultaneously measured both the velocity $\mathbf{v}$ and the nematic director $\mathbf{n}$ fields (SI Section 5).
\begin{figure*}[h!]
	\centering
	\includegraphics[width=1\columnwidth]{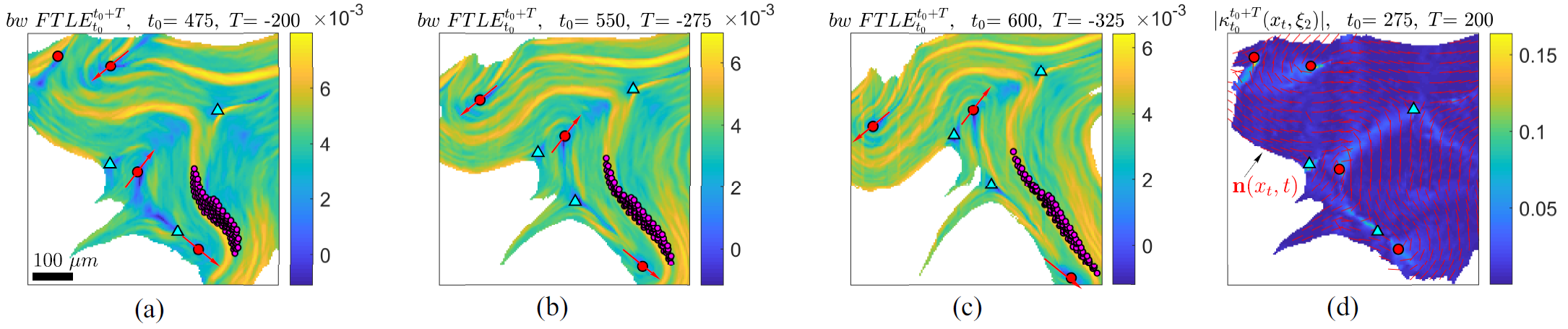} 
	\caption{Active suspension of microtubule bundles and kinesin at the water oil interface with a 2$\mu M$ ATP concentration. (a) $_b\Lambda$ for $\vert T\vert = 200s$ along with the position of nematogens (magenta) attracted to a $_b\Lambda$ ridge and initially released from a circular blob. (b-c) Same as (a) for larger $\vert T\vert$. Red arrows illustrate that +1/2 defects pull the attracting $_b\Lambda$ ridges that in turn  shape the Lagrangian motion of nematogens. (d) Absolute Folding field $\vert \kappa_{t_{0}}^t(\mathbf{x}_t,\mathbf{\xi}_2)\vert$ for $T= 200s$, along with topological defects and the director field at the current time $t_0+T$. Defects are invariably located at regions of high folding-type and low stretching or shearing-type Lagrangian deformation. Time is in s, the colorbar in d encoding Lagrangian folding is in $1/{\mu m}$ while those encoding attraction rates in s$^{-1}$. \href{https://www.dropbox.com/s/iqbiudjwy8gt5h7/Apr_20_FTLEMovieNemato_T275_600_FocReg_v3_HRES.mp4?dl=0}{Movie3} shows the time evolution of $_f\Lambda$ and $_b\Lambda$, along with particle motion.}
	\label{fig:Exp_FoldFTLEDirect}
\end{figure*}
We mark $+1/2$ disclinations with red dots and $-1/2$ disclinations with cyan triangles, as shown in Fig. \ref{fig:FTLEScheme}. Figures \ref{fig:Exp_FoldFTLEDirect}a-c show $_b\Lambda$ for increasing time intervals, along with nematogens (magenta), initially released from a circular blob and \black{eventually} attracted to a $_b\Lambda$ ridge. 
\href{https://www.dropbox.com/s/iqbiudjwy8gt5h7/Apr_20_FTLEMovieNemato_T275_600_FocReg_v3_HRES.mp4?dl=0}{Movie3} shows the time evolution of $_f\Lambda$ and $_b\Lambda$ along with particle positions. The FTLE fields again uncover the organizers of fluid motion, while remaining hidden to trajectory plots. 

Along with $_b\Lambda$, Figs. \ref{fig:Exp_FoldFTLEDirect}a-c show the evolution of topological defects, with red arrows indicating the \black{direction} of motion of positive disclinations. These panels suggest that positive disinclination move and deform $_b\Lambda$ ridges, which, in turns, \black{direct}  particle motion. Interestingly, $+1/2$ defects appear to be in regions of low Lagrangian stretching or shearing deformation as quantified by the FTLE field. By contrast, the Lagrangian folding measure is maximum at defects, as shown in Fig. \ref{fig:Exp_FoldFTLEDirect}d, \black{where} the absolute folding field $\vert \kappa_{t_{0}}^t(\mathbf{x}_t,\bm{\xi}_2)\vert$ \black{is superimposed to} the nematic director $\mathbf{n}$ at the same of time of panel a. 

To quantify our observations on deformations at defects and the correlation of their dynamics with $_b\Lambda$, we turn to numerical simulations of an incompressible ($\mathbf{\nabla}\cdot\mathbf{v} = 0$) planar uniaxial active nematic liquid crystal. We solve the nematodynamic equations 
\begin{subequations}\label{eq:Nematodynamic}
	\begin{gather}
	\rho\frac{D{\bf v}}{Dt}=-\mathbf{\nabla}p + \eta\mathbf{\nabla}^2\mathbf{v}+\mathbf{\nabla}\cdot(\mathbf{\sigma}^e + \mathbf{\sigma}^a)\;,\label{eq:Momentum}\\
	\frac{D{\bf Q}}{Dt}=\lambda S \mathbf{D}+\mathbf{Q}\mathbf{\Omega}-\mathbf{\Omega}\mathbf{Q}+\gamma^{-1}\mathbf{H}\;,\label{eq:Nematic}
	\end{gather}
\end{subequations}
which can be derived from phenomenological arguments or microscopic models, and capture typical experimental statistics (see e.g. Ref. \cite{Giomi2015}). Here, $\rho$ and $\eta$ denote the density and viscosity of the nematic fluid, $D/Dt=\partial_t + \mathbf{v}\cdot\mathbf{\nabla}$ is the material derivative, $\lambda$ is the flow alignment parameter and $\gamma$ is the rotational viscosity \cite{gennes1993physics}. In Eq. \eqref{eq:Nematic}, $\mathbf{Q} = S(\mathbf{n}\mathbf{n}-\mathbf{I}/2)$ denotes the nematic tensor, $0\le S \le 1$ the nematic order parameter, $\mathbf{I}$ the identity tensor, $\mathbf{D}=[\mathbf{\nabla v}+(\mathbf{\nabla v})^{\top}]/2$ is the symmetric and $\mathbf{\Omega}=[\mathbf{\nabla v}-(\mathbf{\nabla v})^{\top}]/2$  the antisymmetric part of the velocity gradient $\mathbf{\nabla v}$. Finally,  \black{$\mathbf{H}=-\delta F/{\delta \mathbf{Q}} =K\nabla^{2}\bm{Q}^{2}-(a_{2}+a_{4}|\bf Q|^{2}){\bf Q}$} is the molecular tensor governing the relaxation dynamics of the nematic phase defined as the variational derivative of the two-dimensional Landau-de Gennes free energy $F=\int f\,dA$ with $f$ the free-energy density \cite{gennes1993physics}
\begin{equation}
\label{eq:LdGEnergy}
f = \frac{1}{2}K|\nabla{\bf Q}|^{2}+\frac{1}{2}a_{2}|{\bf Q}|^{2} + \frac{1}{4}a_{4}|{\bf Q}|^{4}\;,
\end{equation}
where $|\cdot|$ denotes the Frobenius norm \black{(i.e. $|{\bf Q}|^{2}=Q_{ij}Q_{ij}$)}, $K$ is the orientational stiffness relating the elastic free energy to spatial inhomogeneities in the \black{configuration of the nematic tensor and $a_{2}$ and $a_{4}$ are bulk moduli}. Finally, $\mathbf{\sigma}^e = -\lambda \black{S} \mathbf{H} + \mathbf{Q}\mathbf{H} - \mathbf{H}\mathbf{Q}$ denotes the elastic stress \black{arising from a departure from the lowest free energy configuration}, and $\mathbf{\sigma}^a = \alpha\mathbf{Q}$ the contractile ($\alpha > 0$) or extensile ($\alpha < 0$) active stress exerted by the active particles along $\mathbf{n}$. 

\black{Since the typical Reynolds number of microtubules/kinesin suspensions varies in the range $10^{-5}-10^{-3}$, depending on the ATP concentration,} we eliminate the convective derivative in Eq. \eqref{eq:Momentum} and \black{numerically integrate} Eq. \eqref{eq:Nematodynamic} \black{using finite differences} on a $128 \times 128$ \black{collocated grid} with periodic boundary conditions. \black{In all our simulations we set the} parameter values \black{as follows:} $\lambda = 0.1$, $K = 1$, $a_{2}=-1$, $a_{4}=2$, $\gamma = 10$, $\alpha = 25$ \black{and $L=5$,} in \black{previously defined re}scaled units. 
This yields the velocity field $\mathbf{v}$ along with the nematic tensor field $\mathbf{Q}$, from which we identify topological defects (SI Section 2).

\begin{figure*}[h!]
	\centering
	\includegraphics[width=1.05\columnwidth]{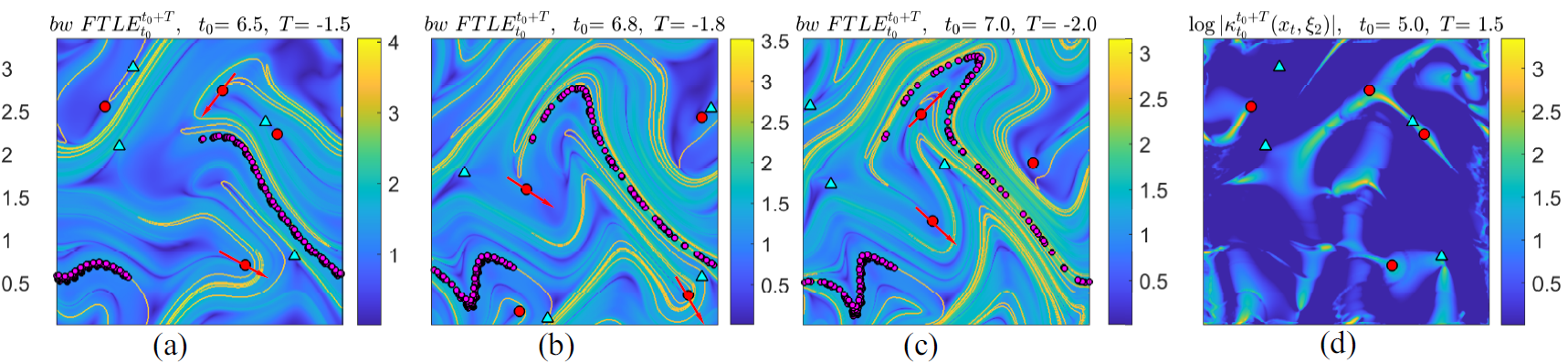}
	\caption{Lagrangian analysis of an active nematic fluid obeying eq. \eqref{eq:Nematodynamic}. 
		(a) $_b\Lambda$ for $\vert T\vert =  1.5$, along with the position of material particles (magenta) attracted to a $_b\Lambda$ ridge and initially released from a circular blob. (b-c) Same as (a) for larger $\vert T\vert$. Red arrows illustrate that +1/2 defects pull the attracting $_b\Lambda$ ridges that, in turn, shape particle motion. (d) Logarithm of the folding field modulus $\vert \kappa_{t_{0}}^t(\mathbf{x}_t,\mathbf{\xi}_2)\vert$ for $T= 1.5$ as in (a). \href{https://www.dropbox.com/s/9qj17xjn6l62al2/Apr_20_Test8Fold_bwFTLE_locInterON_T2.mp4?dl=0}{Movie4} shows the evolution of panels a,d for different $T$.  Positive defects are located at regions of high folding-type and low stretching or shearing-type Lagrangian deformation.}
	\label{fig:BendingFTLENematoTest8Num}
\end{figure*}

Figures \ref{fig:BendingFTLENematoTest8Num}a-c show $_b\Lambda$ for different time intervals $\vert T\vert$, along with the position of an initially circular set of particles (magenta) that are attracted to the a $_b\Lambda$ ridge. 
Analogous to the analysis of the experiments, we find that $_b\Lambda$ ridges are pulled (red arrows) and shaped by moving Eulerian +1/2 disclinations while remaining insensitive to -1/2 disclinations (Figs. \ref{fig:BendingFTLENematoTest8Num}a-c). 
In the SI Section 2, we quantify the correlation between the evolution of $_b\Lambda$ and defects motion. We first find a velocity field that transports and deforms $_b\Lambda$ over increasing $T$, and evaluate it at disclinations. The $_b\Lambda$ evolution along with disclination velocities is available as \href{https://www.dropbox.com/s/idru8ofudmbpdzg/Apr_20_Test8_T3_BWFTLE_DisclVel_v7_dT005_VF.mp4?dl=0}{Movie5}. We then compute the relative angle between the $_b\Lambda$ velocity at defects and the disclination velocities. The mean and standard deviation of the relative angle associated with positive disclinations are seven and five times smaller compared those related to negative disclinations (SI Section 2). 

We find this result nontrivial as $_b\Lambda$ is Lagrangian, i.e., contains information of particle trajectories, while disclinations are Eulerian, hence agnostic to particle paths. This connection could provide a quantitative framework to control the Lagrangian motion, mixing, and deformation of active nematics by steering the position of Eulerian disclinations. Finally, in the SI Section 3, we provide also an aggregate measure of positional coherence by using the inverse of the broadly used $H^{-1}$ mixing norm \cite{Doering2006}, and show how it decreases with increasing $T$ and activity $\alpha$. 


Figure \ref{fig:BendingFTLENematoTest8Num}d shows $\kappa_{t_{0}}^t(\mathbf{x}_t,\bm{\xi}_2)$ associated with the $\vert T\vert =1.5$, as in \black{Fig. \ref{fig:BendingFTLENematoTest8Num}a}. \black{Consistently with our experimental results (Fig. \ref{fig:Exp_FoldFTLEDirect})} 
defects are \black{preferentially} located in regions of high Lagrangian folding and low stretching. We obtain results similar to Fig. \ref{fig:BendingFTLENematoTest8Num} for extensile ($\alpha<0$) active nematics, simulated using the same parameters listed above and $\alpha = -25$. \href{https://www.dropbox.com/s/86y3ppaaaq0wjaj/Apr_20_Test13Fold_bwFTLE_1locInterON_T2.mp4?dl=0}{Movie6} shows the same as \href{https://www.dropbox.com/s/9qj17xjn6l62al2/Apr_20_Test8Fold_bwFTLE_locInterON_T2.mp4?dl=0}{Movie4} for the extensile case. 

\subsection*{Stress gradients are maximal at positive defects}
\begin{figure*}[h!]
	\centering
	\includegraphics[width=1\columnwidth]{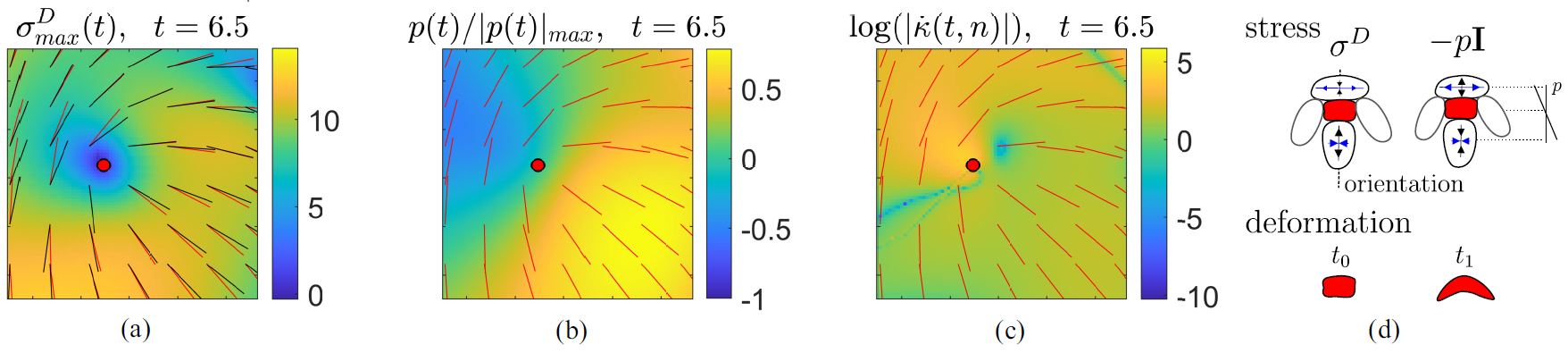}
	\caption{Stress and deformation around an isolated $+1/2$ defect (bottom right in Figs. \ref{fig:BendingFTLENematoTest8Num}a,d) in contractile active nematics. (a) Maximum eigenvalue of the total deviatoric stress $\mathbf{\sigma}^D$ in the proximity of a $+1/2$ defect (red cirlce). The leading eigenvector of $\mathbf{\sigma}^D$ is marked by black lines and the the nematic director field by red lines (see Figs. \ref{fig:StressContribPress} for the separate viscous, elastic and active stress contributions). (b) Pressure field normalized by the spatial maximum pressure in absolute value completely characterizes the isotropic stress $\mathbf{\sigma}^I = -p\mathbf{I}$. (c) Logarithm of the folding rate modulus of the active nematic, computed from eq. \eqref{eq:ThmKappaDotnMT}. (d) Top: Sketch of the deviatoric and isotropic stress distribution near an isolated $+1/2$ defect in contractile active nematics. The arrow size is proportional to the stress level, and blue marks the direction perpendicular to the defect orientation. Bottom: sketch of the material deformation in correspondence the $+1/2$ defect. Figures \ref{fig:StressContribPressExt} and \ref{fig:StressDefPosDefExtens} show equivalent analysis for the extensile case.}
	\label{fig:StressAtDefects}
\end{figure*}

Motivated by the striking deformations associated with  +1/2 defects, we analyze the stress distribution using the simulation data (Fig. \ref{fig:StressAtDefects}). 
We find that both the deviatoric and isotropic total stresses are minimal at +1/2 defects, but have high stress gradients along the defect orientation (\ref{fig:StressAtDefects}a,b), which induces folding deformation as sketched in Fig. \ref{fig:StressAtDefects}d. We obtain similar results for extensile active nematics (SFig. \ref{fig:StressContribPressExt}), where the folding direction is towards the tail of the defect (SFig. \ref{fig:StressDefPosDefExtens}) as opposed to the head (Fig. \ref{fig:StressAtDefects}d). 

{To bridge the gap between the Lagrangian deformations, which accounts for the motion history of the nematic continuum, and Eulerian stresses based on an instantaneous configuration, we derived an exact formula for the Eulerian folding rate}
\begin{equation}
\dot{\kappa}(t,\mathbf{x},\mathbf{n}) = [(\mathbf{\nabla D}(\mathbf{x},t)\mathbf{n})\cdot\mathbf{n}]\cdot\mathbf{n_{\perp}}-\frac{ \mathbf{\nabla} \omega(\mathbf{x},t) \cdot \mathbf{n}}{2}
\label{eq:ThmKappaDotnMT}
\end{equation}
experienced by an infinitesimal patch of nematic fluid with orientation $\mathbf{n}$ (SI Section 1). The folding rate can be computed from $\mathbf{v},\ \mathbf{n}$, and arises from spatial heterogeneities of the rate-of-strain tensor $\mathbf{D}$ and the vorticity $\omega$. If the nematic continuum is an epithelium, for instance, $\dot{\kappa}(t,\mathbf{x},\mathbf{n})$ measures the bending rate experienced by the cell located at $\mathbf{x}$. Using eq. \eqref{eq:ThmKappaDotnMT}, we also find that   $\dot{\kappa}(t,\mathbf{x},\mathbf{n})$ is maximum in the vicinity of +1/2 defects (Fig. \ref{fig:StressAtDefects}c), consistent with the corresponding stress distribution (Figs. \ref{fig:StressAtDefects}a,b) and Lagrangian folding (Fig. \ref{fig:BendingFTLENematoTest8Num}d). 

It is interesting to compare these results with experimental findings that suggest a biological functionality of topological defects in a biological epithelial layer  \cite{Doostmohammadi2018,Saw2017,Kawaguchi2017}. The quantification of deformations and stress at the defects is a key step in elucidating how mechanical stimuli are converted into downstream biochemical signals.
Positive defects with strength $+1/2$ in monolayers of MDCK cells, for example, have been associated with sites of cell apoptosis \cite{Saw2017}, with a possible explanation being high compressive stress at the defect location. This hypothesis has been tested by correlating the isotropic stress averaged over several ($6$) cell sizes in the neighborhood of a topological defect during apoptosis (Fig. 3b in \black{Ref.}~\cite{Saw2017}). 
Our findings, however, show that bending deformations are dominant at positive disclinations, suggesting there can be new relevant mechanisms at play at topological defects in epithelia. 

We observe a clear similarity of the extensile active nematic stress distribution in SFig. \ref{fig:StressContribPressExt}h, with that experimentally measured in monolayers of MDCK cells (Fig. 3a in \black{Ref.}~\cite{Saw2017}) during apoptosis. The peculiar stress and deformation distribution around +1/2 defects, together with the cells' ability to sensing curvature changes \cite{dreher2016snapshot}, may lead to uncovering novel feedback mechanisms in epithelial dynamics. A natural next step is to correlate folding deformation maps in epithelia with the corresponding YAP (Yes-associated protein) distributions \cite{aragona2013mechanical}. 

\subsection*{Discussion}
By combining concepts from nonlinear dynamics, experiments of two-dimensional active nematics and \black{active hydrodynamics} simulations, we found that the motion of active nematics is organized by hidden dynamic (time-dependent) attracting and repelling CSs, whose motion is coupled to that of $+1/2$ topological defects. As they move deform attracting CSs, which in turn \black{regulate collective} motion. Furthermore, the Lagrangian time-scale dependent maps of stretching- and folding-type deformations of a nematic continuum show that $+1/2$ defects occur at locations of high bending and low stretching or shearing type deformations. Motivated by this finding, we have discovered a characteristic stress distribution around +1/2 defects: the stress is minimal at the defect, but its large gradient along the defect's orientation causes differential stress that induces bending.  The bending is towards the defect head (tail) for contractile (extensile) extensile active nematics. Similar stress distributions were measured experimentally in monolayers of MDCK (Madin Darby canine kidney) cells \cite{Saw2017}. 

More broadly, using only the measured velocity and nematic director, our results provide a quantitative framework for assessing the motion, mixing and deformation of active nematics. Emerging experimental evidence associated biological functionality with topological defects of cells orientation \cite{Doostmohammadi2018}, actin fibers orientation \cite{maroudas2020topological}, and the ability of cells to sense and react to bending and stretching deformations \cite{dreher2016snapshot}. From this perspective, our approach quantifies the stretching and folding deformations in a nematic continuum as time-scale dependent maps {with the ability to predict and perhaps eventually control the flows}. Investigating the correlation between the curvature and stretching deformation maps in epithelial layers and other similar systems, could elucidate how cells couple mechanical inputs to intracellular signals in oriented active matter systems.


\section*{Acknowledgements}
\black{We are grateful to Suraj Shankar and Nicola Molinari for helpful discussions. This work is partially supported by} the Schmidt Science Fellowship and the Postdoc Mobility Fellowship from the Swiss National Foundation (M.S.), \black{the Netherlands Organization for Scientific Research (NWO/OCW) as part of the Frontiers of Nanoscience program and the Vidi scheme (L.G.), and the Department of Energy, Office of Basic Energy Sciences under Award No. DESC0019733 (ZD), and the NSF Simons Center for Mathematical and Statistical Analysis of Biology Award No.  1764269 (LM).} 

\onecolumn
\clearpage
\begin{center}
	\textbf{\Large Supplemental Information}
\end{center}

\setcounter{figure}{0}
\setcounter{section}{0}
\setcounter{equation}{0}
\renewcommand{\thefigure}{S\arabic{figure}}

\section*{S1. Lagrangian folding and Eulerian folding rates}
The curvature of an infinitesimal material fiber at time $t$ starting 	from the initial position $\mathbf{x}_0$, with an orientation $\theta$, and curvature $\kappa_0$, due to the transport and deformation induced by the flow map $\mathbf{F}_{t_0}^t(\mathbf{x}_0)$, can be computed \cite{Serra2017Separation} as 
\begin{equation}
\begin{aligned}
\label{eq:Curvaturegeneral}
\kappa_{t_0}^t:\mathbb{R}^2\times\mathbb{S}^1\times\mathbb{R}\rightarrow &\  \mathbb{R},\\
\kappa_{t_0}^t(\mathbf{x}_0,\theta,\kappa_0)=& 	\frac{[(\mathbf{\nabla^2 F}_{t_0}^t(\mathbf{x}_0)\mathbf{e}_{\theta})\cdot\mathbf{e}_{\theta}]\cdot [\mathbf{\nabla_{\perp} F}_{t_0}^t(\mathbf{x}_0)\cdot\mathbf{e}_{\theta}]}{( \mathbf{e}_{\theta}\cdot [ \mathbf{C}_{t_0}^t(\mathbf{x}_0)\cdot\mathbf{e}_{\theta}])^{3/2}}+\kappa_0 \frac{\det[\mathbf{\nabla F}_{t_{0}}^t(\mathbf{x}_0)]}{( \mathbf{e}_{\theta}\cdot [ \mathbf{C}_{t_0}^t(\mathbf{x}_0)\cdot\mathbf{e}_{\theta}])^{3/2}},\\
\end{aligned}
\end{equation}
where $\mathbf{e}_{\theta} = [\cos\theta,\sin\theta]$ and $(\nabla^2 F_{t_{0}}^t(\mathbf{x}_0)e_{\theta_k})_{ij}=\tiny{\sum\limits_{k}}{F_{t_{0}}^t}_{i,jk}(\mathbf{x}_0)e_{\theta_k},\ i,j,k\in\{1,2\}$. By evaluating $\kappa_{t_0}^t(\mathbf{x}_0,\theta,\kappa_0)$ for an initially straight ($\kappa_0=0$) fiber aligned with the dominant eigenvector of $\mathbf{C}_{t_0}^t$ ($\mathbf{e}_{\theta} = \mathbf{\xi}_{2}$), we obtain Eq. (4) in the main text. 

In the instantaneous limit ($t=t_0$), the material curvature rate of the material fiber \cite{Serra2017Separation} is given by 
\begin{equation}
\begin{aligned}
\frac{d\kappa_{t_0}^t(\mathbf{x}_0,\theta,\kappa_0)}{dt}\vert_{t=t_0} &= {\dot{\kappa}}_{t_{0}}(\mathbf{x}_0,\theta,\kappa_0)\\
&= [(\mathbf{\nabla D}(\mathbf{x}_0,t_0)\mathbf{e}_\theta)\cdot\mathbf{e}_\theta]\cdot{\mathbf{e}_\theta}_{\perp}-\frac{\mathbf{\nabla} \omega(\mathbf{x}_0,t_0)\cdot \mathbf{e}_\theta}{2} +\kappa_0\bigg{[}\mathbf{\nabla}\cdot \mathbf{v}(\mathbf{x}_0,t_0)-3 \mathbf{e}_\theta\cdot [\mathbf{D}(\mathbf{x}_0,t_0)\cdot\mathbf{e}_\theta]\bigg{]},
\label{eq:ThmKappaDot}
\end{aligned}
\end{equation}
where $\mathbf{D}$ denotes the rate-of-strain tensor, $\omega$ the vorticity, $\mathbf{\nabla}\cdot \mathbf{v}$ the divergence of the flow and $(\nabla D(\mathbf{x}_0,t_0)e_{\theta_k})_{ij}=\tiny{\sum\limits_{k}}D_{ij,k}(\mathbf{x}_0)e_{\theta_k},\ i,j,k\in\{1,2\}$. Integrating Eq. \eqref{eq:ThmKappaDot} along trajectories $\mathbf{F}_{t_0}^t(\mathbf{x}_0)$, provides an alternative formula to compute Eq. \eqref{eq:Curvaturegeneral} from know Eulerian quantities (see Eq. B1 in \cite{Serra2017Separation}). 
Evaluating eq. \eqref{eq:ThmKappaDot} along the current nematic director ($\mathbf{e}_\theta  \equiv \mathbf{n}$), and assuming $\kappa_{0}\equiv 0$, one can compute the instantaneous folding rate experienced by a nematic continuum using only the velocity and the director fields inputs as
\begin{equation}
\dot{\kappa}(t,\mathbf{x},\mathbf{n}) = [(\mathbf{\nabla D}(\mathbf{x},t)\mathbf{n})\cdot\mathbf{n}]\cdot\mathbf{n_{\perp}}-\frac{ \mathbf{\nabla} \omega(\mathbf{x},t) \cdot \mathbf{n}}{2}.
\label{eq:ThmKappaDotn}
\end{equation}
We note that eq. \eqref{eq:ThmKappaDotn} allows quantifying the folding rate contribution coming from spatial inhomogeneities of the rate of strain tensor and the vorticity. To deploy these results in an experimental setting, consider a nematic continuum that describes an epithelial tissue where $\mathbf{n}$ represents the cell orientation field. Then, for instance, eq. \eqref{eq:ThmKappaDotn} quantifies the instantaneous bending rate of epithelial cells assuming that cells have initially zero curvature. 

\section*{S2. Positive defects move and deform attracting Lagrangian coherent structures}
We quantify the influence of moving defects on the motion and deformation of attracting Lagrangian Coherent Structures identified with the backward FTLE. Denoting by $\mathbf{q}(\cdot,t) : = [Q_{11}(\cdot,t),Q_{12}(\cdot,t)]^\top$ the vector containing the independent entries of the nematic tensor, and by $\mathbf{x^d}(t):=\underset{\mathbf{x^d}(t)}{\arg}\mathbf{q}(\mathbf{x^d}(t),t)=0$ the time-$t$ position of disinclinations, we compute the disclination velocities by Taylor-expanding the equation defining defect locations
\begin{equation}
\label{eq:Discl}
\underset{\mathbf{x^d}(t+\delta t)}{\arg}{\mathbf{q}(\mathbf{x^d}(t+\delta t),t+\delta t)} \approx \underset{\mathbf{x^d}(t)}{\arg}\mathbf{q}(\mathbf{x^d}(t),t) + [\mathbf{\nabla}\mathbf{q}(\mathbf{x^d}(t),t)\mathbf{v^d}(t)+\partial_t\mathbf{q}(\mathbf{x^d}(t),t)]\delta t =0,
\end{equation}
and requiring the leading order term to vanish, i.e.
\begin{equation}
\label{eq:Discl2}
\mathbf{v^d}(t)= - [\mathbf{\nabla}\mathbf{q}(\mathbf{x^d}(t),t)]^{-1}\partial_t\mathbf{q}(\mathbf{x^d}(t),t).
\end{equation}
By the implicit function theorem, $\mathbf{v^d}(t)$ exists whenever $[\mathbf{\nabla}\mathbf{q}(\mathbf{x^d}(t),t)]$ is invertible. Following \cite{tricoche2004topology}, we compute the index of the disclination at $\mathbf{x^d}$ as 
\begin{equation}
\label{eq:DisclIdx}
\text{ind}_{\text{d}} = \frac{1}{2\pi}\sum_{i=1}^{n}\Delta_i,\quad \Delta_i = \phi_{i+1}-\phi_{i} -\pi\ \text{round}\bigg(\frac{\phi_{i+1}-\phi_{i}}{\pi}\bigg),\ \phi_{n+1} = \phi_{1},
\end{equation}
where $\phi_i,\ i=1,...,4$ denotes the angle between the nematic director and the horizontal axis at each of the grid points surrounding the defect.

\begin{figure*}[h!]
	\centering
	\subfloat[]{\includegraphics[height=.3\columnwidth]{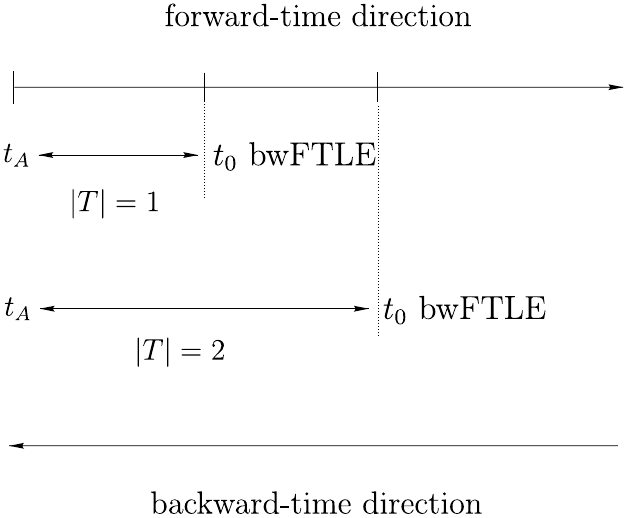}\label{fig:bwFTLEtiming}}	\hfill 
	\subfloat[]{\includegraphics[height=.3\columnwidth]{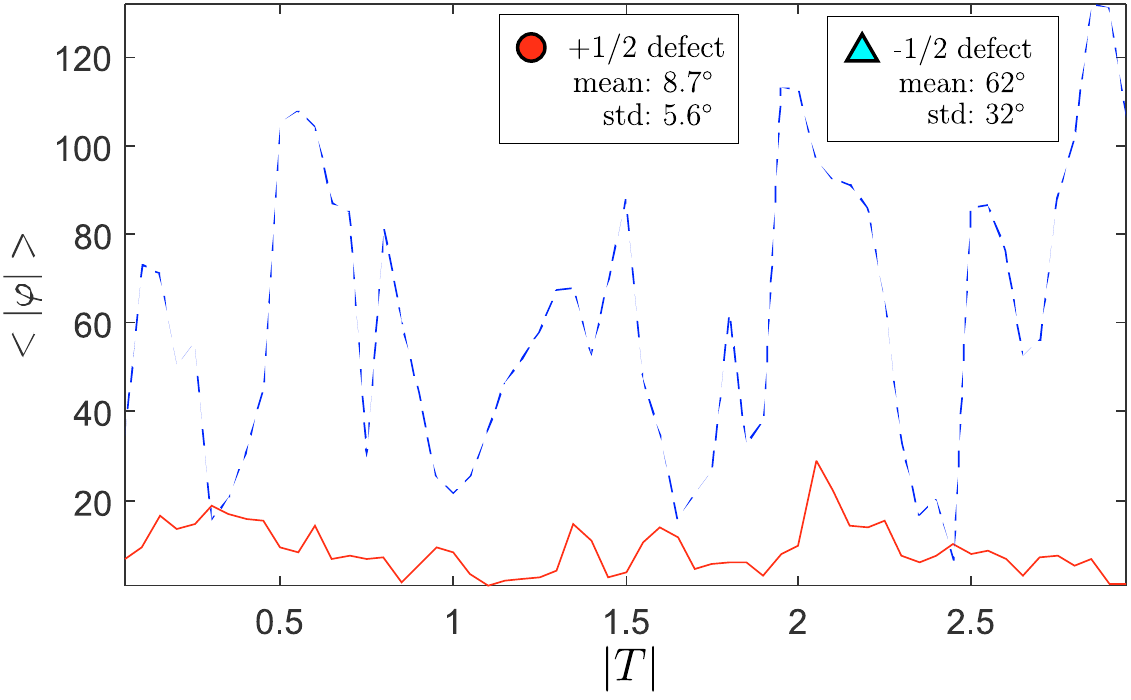}\label{fig:bwFTLEDisclvelAllign}}
	\caption{(a) Time intervals for backward-time FTLE computations. (b) Average angle, in degrees, between a vector field describing the deformation of the bw FTLE, and disclination velocities as a function of $\vert T \vert$. The bw FTLE evolution along with disclination velocities is available as \href{https://www.dropbox.com/s/idru8ofudmbpdzg/Apr_20_Test8_T3_BWFTLE_DisclVel_v7_dT005_VF.mp4?dl=0}{Movie5}.}
	\label{fig:AllignbwFTLEvel_disclVel}
\end{figure*}

To quantify the deformation of the FTLE field with respect to $\vert T\vert$, we use Digital Particle Image Velocimetry to obtain a fictitious velocity field, $\mathbf{v}_{FTLE}(\mathbf{x},T)$, that deforms the $\text{FTLE}_{t_{0}}^{t_{0}-\vert T\vert}$ onto $\text{FTLE}_{t_{0+\Delta T}}^{t_{0}-(\vert T\vert + \Delta T)}$ (Fig. \ref{fig:bwFTLEtiming}). 
Evaluating such velocity field at the current location $\mathbf{x^d}_i(T)$ of defect $i$, we compute the relative angle between $\mathbf{v}_{FTLE}$ and $\mathbf{v^d}$ as 
\begin{equation}
\label{eq:relangle}
\mathbf{\varphi}_i(T)= \arccos{\frac{ \mathbf{v}_{FTLE}(\mathbf{x^d}_i(T),T)\cdot \mathbf{v^d}_i(T)}{\vert \mathbf{v^d}_i(T) \vert \vert \mathbf{v}_{FTLE}(\mathbf{x^d}_i(T),T)\vert }}.
\end{equation}
We note that $T$ automatically specify $t_0$ and vice-versa because the initial time of our analysis $t_A$ is fixed. Fig. \ref{fig:bwFTLEDisclvelAllign} shows the average angle $<\vert\varphi\vert>$ between $\mathbf{v}_{FTLE}$ and positive and negative disinclinations at each $\vert T \vert$, quantitatively confirming that positive disinclinations move in directions similar to $\mathbf{v}_{FTLE}$ compared to negative disinclinations. The overall angular distance between positive disinclination and $\mathbf{v}_{FTLE}$ has a mean of $8.7^\circ$. By contrast, for negative disinclinations the average misalignment is $62^\circ$. 

\section*{S3. Positional coherence as a function of time and activity}

\begin{figure*}[h!]
	\centering
	\hfill
	\subfloat[]{\includegraphics[height=.3\columnwidth]{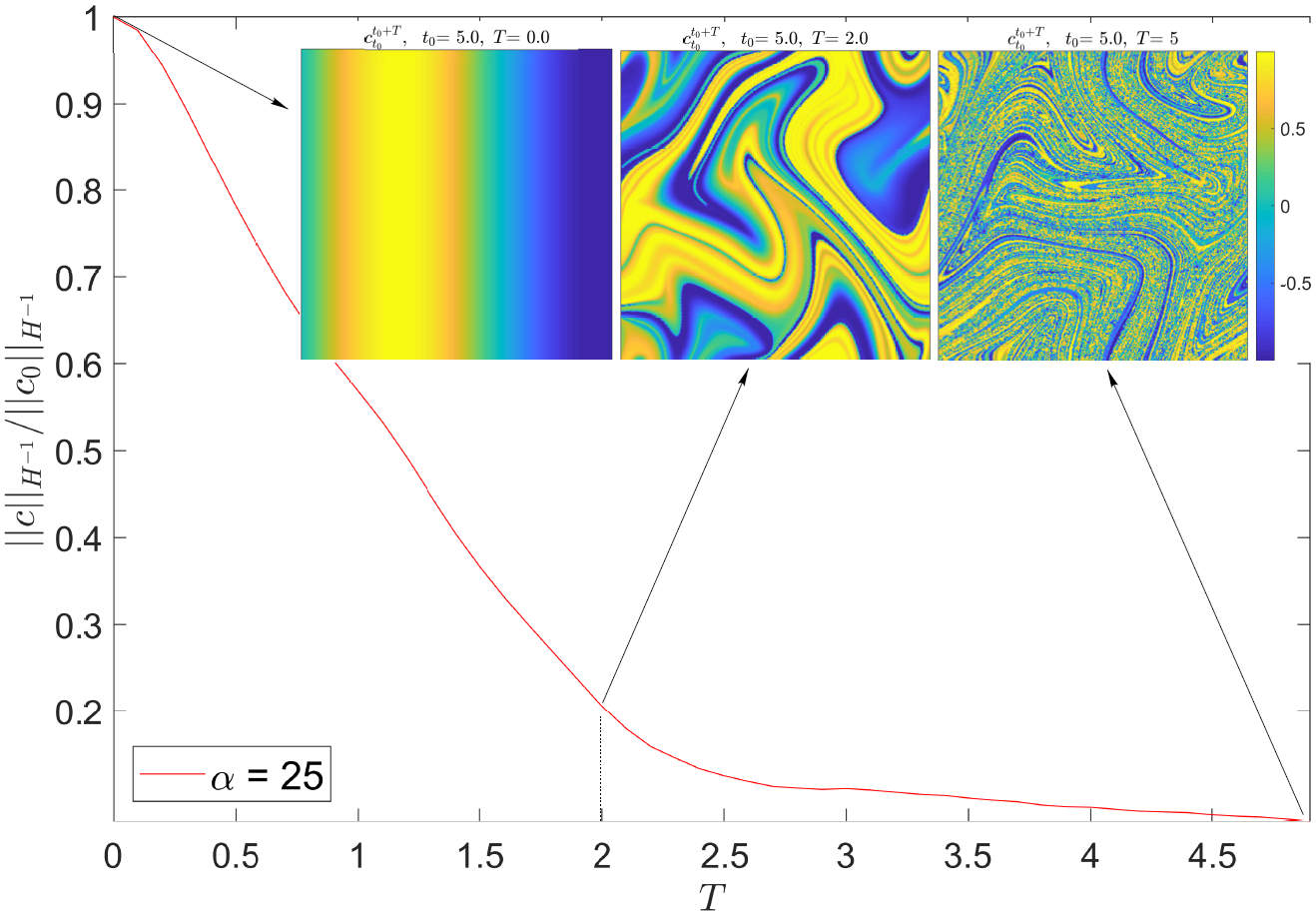}\label{fig:MixAlpha25}}
	\hfill 
	\subfloat[]{\includegraphics[height=.3\columnwidth]{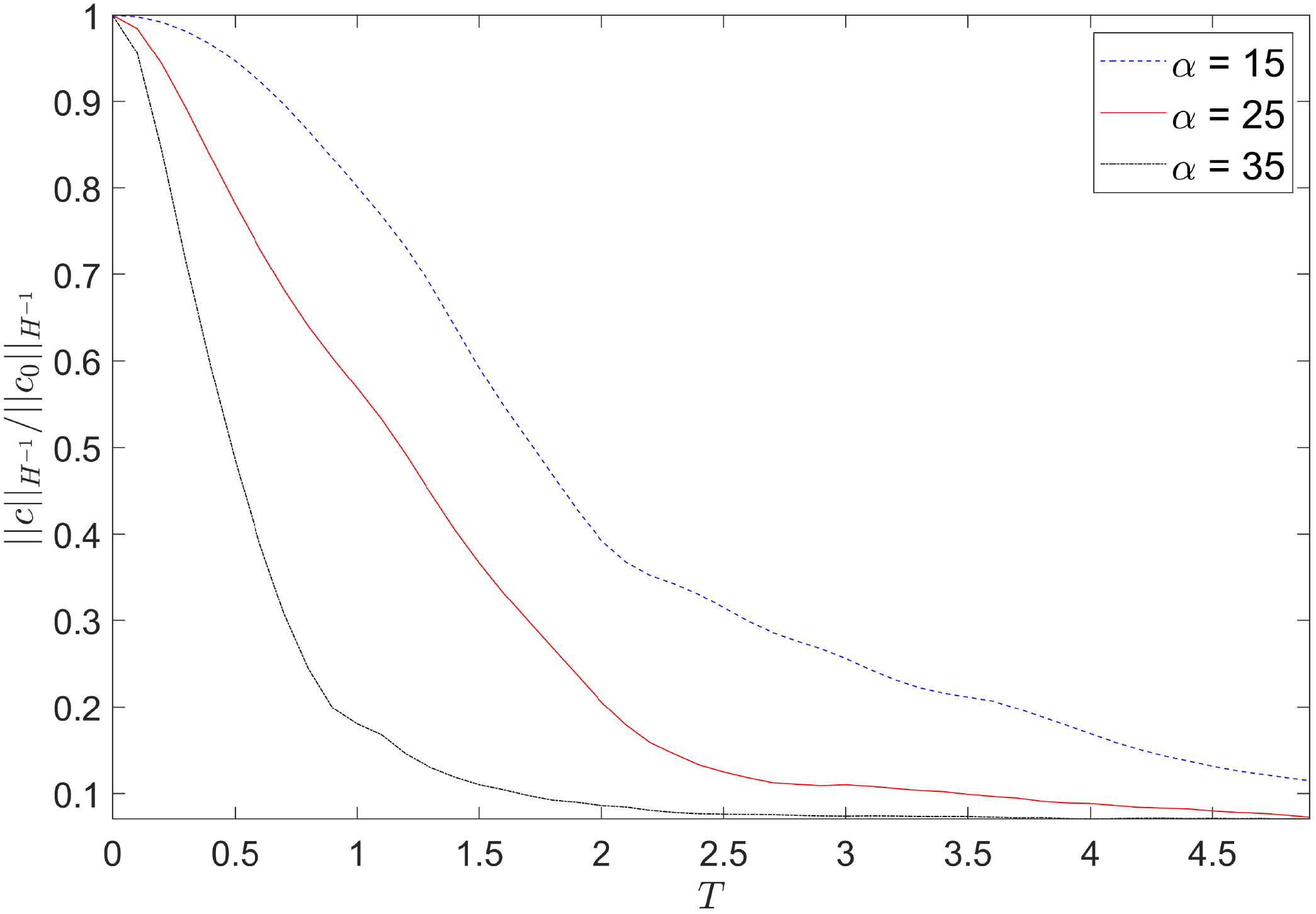}\label{fig:MixVAriousAlpha}}
	\hfill 	\hfill
	\caption{(a) $H^{-1}$ mixing norm normalized by its initial value for the activity value $\alpha=25$. The insets show the concentration fields at different times computed solving Eq. \eqref{eq:PAssiveAdvection} with initial distribution $c_0(\mathbf{x}) = \sin x$. The complete time evolution of the concentration field is available as \href{https://www.dropbox.com/s/rl3vaz0o066ccdq/Test8_T5_PassScalarType5.mp4?dl=0}{Movie7}. (b) Same as (a) for different activity values.}
	\label{fig:Mixing}
\end{figure*}
We quantify how coherence varies with the time scale $T$ and the activity parameter $\alpha$, by studying the stirring exerted by an incompressible active nematic flow on a passive scalar $c(\mathbf{x},t)$. The passive scalar evolves according to 
\begin{equation}
\label{eq:PAssiveAdvection}
\partial_t c + \mathbf{v}\cdot \mathbf{\nabla}c = 0
\end{equation}
with the initial condition $c(\mathbf{x},t_0) = c_0(\mathbf{x})$.
To measure stirring, we use the $H^{-1}$ mixing norm, broadly adopted in fluid flows \cite{Doering2006}, and defined as 
\begin{equation}
\begin{aligned}
\label{eq:MixingNorm}
\vert\vert c(\cdot,t) \vert\vert^2_{H^{-1}} = \vert\vert \vert \mathbf{\nabla}\vert^{-1}c(\cdot,t)\vert\vert^2_{L^{2}} = \sum_{\mathbf{k}\neq0}\vert\mathbf{k}\vert^{-2}\vert {\hat{c}}_{\mathbf{k}}(t)\vert^2,
\end{aligned}
\end{equation}
where 
\begin{equation}
\label{eq:FFT}
\hat{c}_{\mathbf{k}}(t) = \frac{1}{L}\int_{[0,L]^2}e^{-i\mathbf{k}\cdot\mathbf{x}}c(\mathbf{x},t)d\mathbf{x}
\end{equation}
are the Fourier coefficients of $c(\mathbf{x},t)$. The $H^{-1}$ measures the variance of a low-pass-filtered image of the concentration field; the smaller it is, the less coherence (more mixed) is the scalar field on large spatial scales. In our analysis, $c_0(\mathbf{x}) = \sin x$.

Fig. \ref{fig:MixAlpha25} shows the $H^{-1}$ mixing norm normalized by its initial value as a function of $T$ for nematic flow analyzed in Fig.\ref{fig:BendingFTLENematoTest8Num}, with activity value $\alpha=25$. The insets show the concentration fields at three different times, while the complete time evolution is available as \href{https://www.dropbox.com/s/rl3vaz0o066ccdq/Test8_T5_PassScalarType5.mp4?dl=0}{Movie7}. Fig. \ref{fig:MixVAriousAlpha} shows the normalized $H^{-1}$ mixing norm for three different values of activity. As expected, spatial coherence decreases with increasing $T$, and higher activity $\alpha$. Finally, we note that the activity values used here are higher than those seen in typical biological nematic systems such as epithelial, fibroblast and stem cells \cite{Doostmohammadi2018}, where positional coherence is present for larger $T$. 

\section*{S4. Stress at +1/2 defects}
\begin{figure*}[h!]
	\centering 
	\includegraphics[height=0.7\columnwidth]{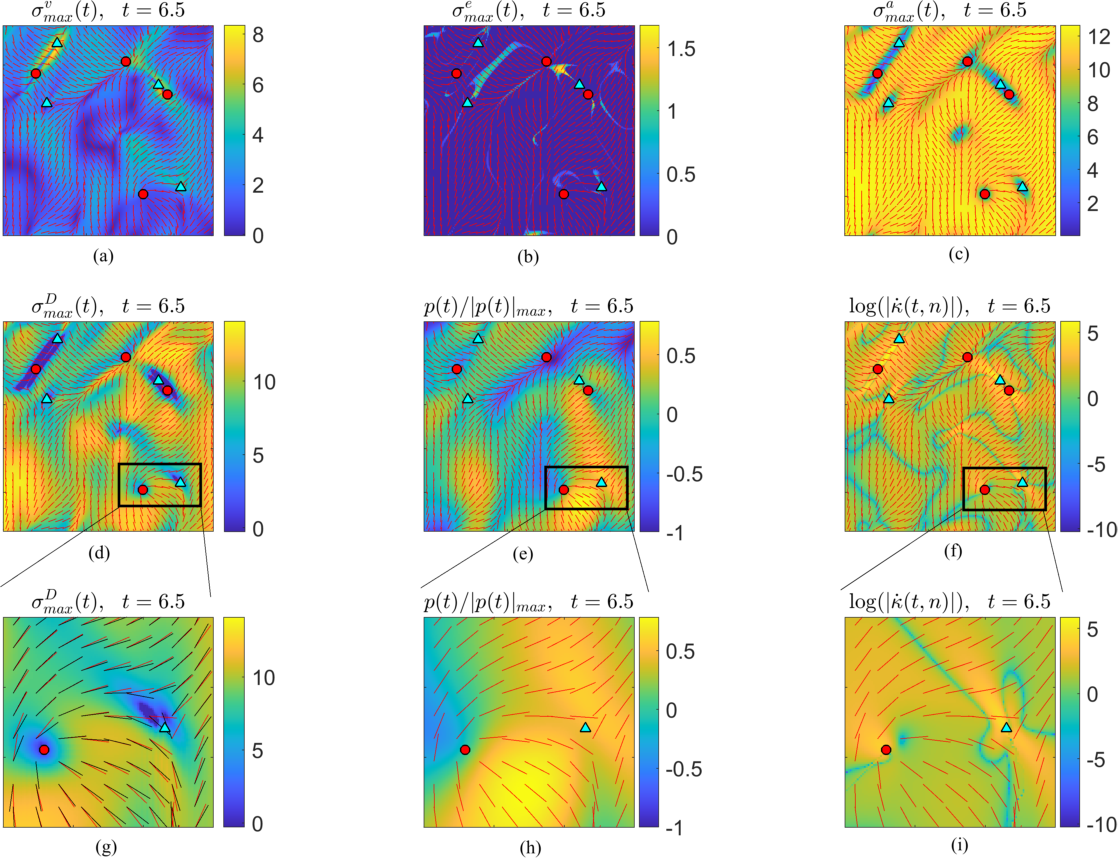}
	\caption{Stress associated with a contractile active nematic ($\alpha>0$). (a-d) Maximum eigenvalue of  $\mathbf{\sigma}^v,\mathbf{\sigma}^e,\mathbf{\sigma}^a,\mathbf{\sigma}^D=\mathbf{\sigma}^v+\mathbf{\sigma}^e+\mathbf{\sigma}^a$, along with the topological +1/2 (cirlces), -1/2 (triangles) and the nematic director field in red. (e) Pressure field normalized by the spatial maximum pressure in absolute value completely characterizes the isotropic stress $\mathbf{\sigma}^I = -p\mathbf{I}$. (f) Logarithm of the folding rate modulus of the active nematic, computed from eq. \eqref{eq:ThmKappaDotn}. (g-i) Zoomed view of the insets in  (d-f). The black direction field in (g) shows the leading eigenvector of $\mathbf{\sigma}^D$.
		The time evolution of the above panels is available as \href{https://www.dropbox.com/s/lf0budybqnhxv9h/Apr_20_Test8_TotalStressesPressKdot_Discl.mp4?dl=0}{Movie8}.}
	\label{fig:StressContribPress}
\end{figure*}

To quantify the correlation between the total stress and the location and dynamics of topological defects, we consider the solution of the neamtodynamic model \eqref{eq:Nematodynamic} analyzed in Fig. \ref{fig:BendingFTLENematoTest8Num}. We first note that both the deviatoric viscous stress $\mathbf{\sigma}^v$  and the elastic stress $\mathbf{\sigma}^e = -\lambda S\mathbf{H} + \mathbf{Q}\mathbf{H} - \mathbf{H}\mathbf{Q}$ are traceless; the former follows from incompressibility, while the latter follows because $\text{tr}[\mathbf{H}]=\partial F_{LdG}/\partial Q_{11}+\partial F_{LdG}/\partial Q_{22}=0,\ Q_{11}=-Q_{22}$ and $\text{tr}[\mathbf{Q}\mathbf{H} - \mathbf{H}\mathbf{Q}]=2\text{tr}[\text{skew}(\mathbf{Q}\mathbf{H})]=0$. Finally, the active stress is also traceless because $\text{tr}[\mathbf{\sigma}^a]=\alpha \text{tr}[\mathbf{Q}]=0$.  Thus, the maximum and minimum eigenvalues of the above stress tensors have equal magnitude and opposite signs, hence providing a scalar representation of both their maximum and minimum stress contributions. Figures \ref{fig:StressContribPress}a-d show the maximum eigenvalue of $\mathbf{\sigma}^v,\mathbf{\sigma}^e,\mathbf{\sigma}^a$ and the total deviatoric stress  $\mathbf{\sigma}^D=\mathbf{\sigma}^v+\mathbf{\sigma}^e+\mathbf{\sigma}^a$, along with the defects location and the nematic director field (red). Surprisingly, and contrary to propositions in the literature \cite{Saw2017}, we find that the maximum total deviatoric stress is minimum at defects (Fig. \ref{fig:StressContribPress}d and Fig. \ref{fig:StressContribPress}g, which shows a zoomed version of the inset in Fig. \ref{fig:StressContribPress}d), meaning that the defect locations are regions subject to minimum shear stress. The black arrows in Fig. \ref{fig:StressContribPress}g represent the leading eigenvector field of $\mathbf{\sigma}^D$. 

\begin{figure*}[h!]
	\centering
	\includegraphics[height=0.7\columnwidth]{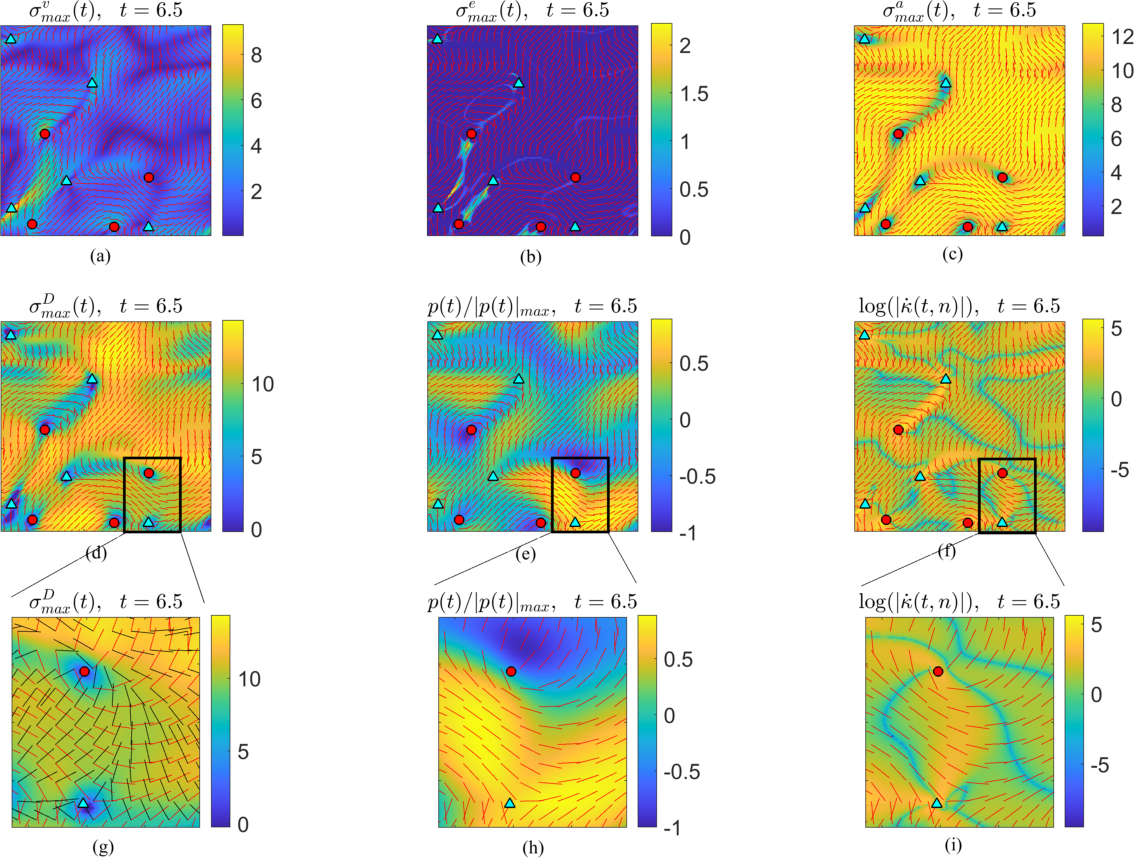}
	\caption{Stresses associated with an extensile active nematic ($\alpha=-25$) following the analysis shown in Fig. \ref{fig:StressContribPress}.
		The time evolution of the above panels is available as \href{https://www.dropbox.com/s/8r40ekbu1r14cs0/Apr_20_Test13_TotalStressesPressKdot_Discl.mp4?dl=0}{Movie9}. Compare panel h with with Fig. 3a in \cite{Saw2017}, which shows the experimentally measured isotropic stress within monolayers of MDCK (Madin Darby canine kidney) cells, in the vicinity of $\pm 1/2$ nematic defects in the cell orientation field.}
	\label{fig:StressContribPressExt}
\end{figure*}

\begin{figure*}[h!]
	\centering
	\hfill 
	\subfloat[]{\includegraphics[height=.35\columnwidth]{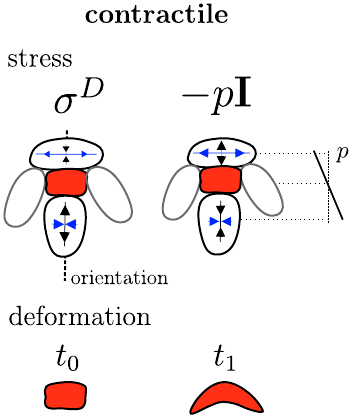}\label{fig:StressDefPosDefContract}}
	\hfill 
	\subfloat[]{\includegraphics[height=.35\columnwidth]{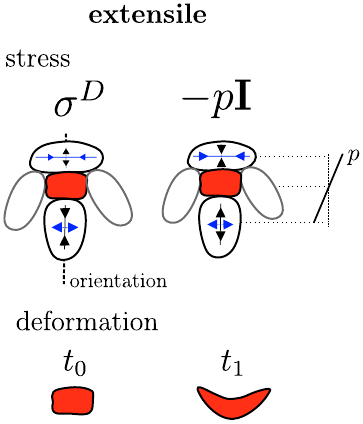}\label{fig:StressDefPosDefExtens}}
	\hfill
    \hfill
	\caption{Schematic of the local stress and deformation around a $+1/2$ defect in contractile (a) and extensile (b) active nematics. (a) Top: Sketch of the deviatoric and isotropic stress distribution (consistent with Figs. \ref{fig:StressContribPress}g,h) near a $+1/2$ defect. The arrow size is proportional to the stress level, and blue marks the direction perpendicular to the defect orientation. Bottom: sketch of the material deformation induced by the stress distribution in neighborhood of a $+1/2$ defect. (b) Same as a for an extensile active nematic ($\alpha<0$). Top: Sketch of the deviatoric and isotropic stress distribution (consistent with Figs. \ref{fig:StressContribPressExt}g,h) near a $+1/2$ defect. Bottom: folding deformation induced by the stress distribution near a $+1/2$ defect.}
	\label{fig:OceanTrap_technical}
\end{figure*}

Because $\mathbf{\sigma}^v,\mathbf{\sigma}^e,\mathbf{\sigma}^a$ are traceless, the only isotropic stress $\mathbf{\sigma}^I = -p\mathbf{I}$ is given by the pressure. With our convention, positive pressure indicates compressing isotropic stress.  Following the numerical scheme in \cite{Giomi2015}, we have solved eq. \eqref{eq:Nematodynamic} by using the streamfunction-vorticity formulation. To recover the pressure, we solve the Poisson equation 
\begin{equation}
\mathbf{\nabla}^2p = \mathbf{\nabla}\cdot[\mathbf{\nabla}\cdot[\mathbf{\sigma}^e + \mathbf{\sigma}^a]],
\label{eq:Poisson}
\end{equation}
obtained by taking the divergence of eq. \eqref{eq:Momentum}, and using incompressibility. Solving eq. \eqref{eq:Poisson}, we determine the pressure distribution up to a constant that will not affect the pressure topology. Here we set this constant such that the spatial average of the pressure is zero. Figures \ref{fig:StressContribPress}e,h show the pressure field normalized by the maximum pressure in absolute value. Interestingly, we find that positive defects are typically located in regions of low isotropic stress. A closer look at Figs. \ref{fig:StressContribPress}g,h, reveals that while deviatoric and isotropic stresses are low at positive defects, the corresponding stress gradients are high. In Fig. \ref{fig:StressDefPosDefContract}, we show that the topology of both the deviatoric and isotropic stresses induce differential stresses (blue), perpendicular to the $+1/2$ defect orientation, that bends the active nematic towards the head of the defect. This peculiar stress distribution leaves a clear kinematic footprint in the deformation of the nematic medium, as shown by the high values of Lagrangian folding at $+1/2$ defects (Fig. \ref{fig:BendingFTLENematoTest8Num}d), as well as the Eulerian folding rate along $\mathbf{n}$, computed from eq. \eqref{eq:ThmKappaDotn} and shown in Figs. \ref{fig:StressContribPress}f,i. 
Figures \ref{fig:StressContribPressExt} and \ref{fig:StressDefPosDefExtens} show the same analysis of Figs. \ref{fig:StressContribPress} and \ref{fig:StressDefPosDefContract} for extensile active nematics, obtained by solving eq. \eqref{eq:Nematodynamic} using the same parameters as the contractile case and $\alpha=-25$.

It is worth pointing out that Fig. \ref{fig:StressContribPressExt}h is   similar to the experimentally measured isotropic stress within monolayers of MDCK (Madin Darby canine kidney) cells, in the vicinity of $\pm 1/2$ nematic defects in the cell orientation field \cite{Saw2017}.  Indeed,  Fig.3a of \cite{Saw2017} shows that $+1/2$ defects are located in a region of zero isotropic stress, and high isotropic stress gradient along the orientation of the defect, precisely as in Fig. \ref{fig:StressContribPressExt}h. As in the contractile case, at $+1/2$ defects the deviatoric stress is zero and has a significant gradient along the defect's orientation (Fig. \ref{fig:StressContribPressExt}g). Fig. \ref{fig:StressDefPosDefExtens} shows the effect of the stress distribution around the $+1/2$ defects, which induces high folding as confirmed both by the Lagrangian finite-time folding (Fig. \ref{fig:BendingFTLENematoTest8Num}d) and the Eulerian folding rate (Fig. \ref{fig:StressContribPressExt}i). The peculiar stress distribution of low stress and high stress gradients around $+1/2$ defects is present in both contractile and extensile active nematics. In the former, the stress gradient induces a folding deformation towards the head of the defect (Fig. \ref{fig:StressDefPosDefContract}), while in the latter towards the tail (Fig. \ref{fig:StressDefPosDefExtens}). 


\section*{S5. Experimental data}

\subsection*{Velocity and Orientation Fields}
We prepared a microtubule-based active nematic at 1.4mM ATP doped with a small fraction of Alexa-647 labeled MTs \cite{decamp2015orientational,VApaper} . We imaged the sample both using LC-PolScope and epifluorscence microscopy. LC-PolScope provides a direct measurement of the orientation field of the MTs \cite{LCPolScope}. Particle image velocimetry was used to find the velocity field from the fluorescence images. The LC-PolScope and fluorescence images were taken sequentially within 2s of each other; however, we treat the lag as negligible in the data as calculating the velocity field coarse grains the data in time. Imaging was done on a Nikon Ti Eclipse equipped with Andor Neo camera.

\subsection*{Photobleaching}

We prepared microtubule-based active nematic at 18$\mu$M ATP with Alexa-647 labeled MTs \cite{decamp2015orientational}.  It was important to use a low amount of ATP to slow the dynamics so that the timescale of bleaching was faster than the movement of the material. We included a small fraction of MTs labeled with Azide-DBCO-488 to simultaneously bleach regions and measure the velocity field of the material. We used a Leica SP8 Confocal with a 20X NA 0.75 air objective to bleach and image the sample. Since $\textrm{Image Brightness} \propto \left(\textrm{NA}^2/M\right)^2$, bleaching is most efficient at low magnification and high NA. To bleach, we decreased the range over which the galvo-mirror scans by 20 times and turned a 633nm laser power to its maximum. With this combination, we were able to bleach in under 5 seconds so that the distortions due to material movement were minimal. Using Leica software, we were able to define regions of interest to bleach defined shapes. To image the sample we reduced the 633nm laser power to 0.5\% of its maximum and simultaneously imaged with a 488nm laser. 

\bibliographystyle{ieeetr}
\bibliography{ReferenceList3}

\end{document}